\documentclass{iau}
\usepackage{graphicx}

\title[Interaction with dense mass loss] 
{Supernova interaction with dense mass loss}

\author[Roger A. Chevalier]   
{Roger A. Chevalier}

\affiliation{Dept. of Astronomy, University of Virginia\\
P.O. Box 400325, Charlottesville, VA 22903, USA \\ email: {\tt rac5x@virginia.edu} \\[\affilskip]
}

\pubyear{2013}
\volume{296}  
\pagerange{119--126}
\jname{Supernova Environmental Impacts}
\editors{A. Ray \& R.A. McCray, eds.}
\begin{document}

\maketitle

\begin{abstract}
Supernovae of Type IIn (narrow line) appear to be explosions that had strong mass loss
before the event, so that the optical luminosity is powered by the circumstellar interaction. 
If the mass loss region has an optical depth $>c/v_s$, where $v_s$ is the shock velocity,
the shock breakout occurs in the mass loss region and a significant fraction of the
explosion energy can be radiated.
The emission from the superluminous SN 2006gy and the normal luminosity SN 2011ht
can plausibly be attributed to shock breakout in a wind, with SN 2011ht being a low
energy event.
Superluminous supernovae of Type I may derive their luminosity from interaction
with a mass loss region of limited extent.
However, the distinctive temperature increase to maximum luminosity has not
been clearly observed in Type I events.
Suggested mechanisms for the strong mass loss include pulsational pair
instability, gravity-waves generated by instabilities in late burning phases, and
binary effects.

\end{abstract}

\firstsection 
\section{Introduction}

Once the radiation from the supernova shock wave emerges from the stellar surface, the interaction with
the surrounding medium commences.
In the case of core collapse supernovae, the initial interaction is with mass loss from the progenitor star.
At low circumstellar densities, the emission from the interaction is a small part of the supernova power and typically involves
X-ray and radio radiation.
This is the case for  Type IIP supernovae, where the inferred circumstellar density is roughly consistent with that expected
from the red supergiant progenitors of the these events (\cite[Chevalier et al. 2006]{chevalier06}).
However, there are supernovae, of Type IIn, that are thought to have most of their peak optical power
come from interaction with a circumstellar medium (e.g., \cite[Chugai 
\& Danziger 1994]{chugai94}).
Here I discuss cases where the circumstellar interaction is an important part of the
optical emission.
The shock breakout phase is treated in Section 2 and the initial viscous shock propagation
in Section 3.
Speculations on the formation of the dense circumstellar medium are discussed in Section 4 and
the final comments in Section 5.

\section{Shock Breakout}

A useful way of considering the enhanced supernova luminosity that can result from very
dense circumstellar interaction is in terms shock breakout in the dense gas
(\cite[Chevalier \& Irwin 2011]{chevalier11}, \cite[Balberg \& Loeb 2011]{balberg2011}).
While a shock front is inside the supernova progenitor, the shock front is
mediated by radiation for standard supernova conditions.
The optical depth across the shock transition for a radiation dominated shock
is $\sim c/v_s$, where $v_s$ is the shock velocity and $c$ is the speed of light.
This result is determined by the condition that the hydrodynamic time for the
shock wave to cross the shock thickness is approximately equal to the diffusion
time for the radiation.
When the shock front reaches a position that is at an optical depth of $\sim c/v_s$ from the surface,
radiation can diffuse out to optically thin layers and shock breakout occurs.
For low circumstellar densities, the timescale for the breakout event is typically quite brief and depends on the radius
of the progenitor star, $R$.
For a red supergiant progenitor, the most extended normal progenitor, the supernova
shock takes about a day to traverse the star and the breakout phase lasts about
half an hour, determined by the light travel time across the star, $R/c$.

A low density, optically thin circumstellar medium is not expected to have much
influence on the optical light near maximum, although it can be very important for
radio and X-ray emission.
At higher circumstellar densities, the medium becomes optically thick and the breakout
radiation must diffuse out through the circumstellar medium.
The result is that the radiated energy in the breakout is not changed, but the radiation
is emitted over a longer period of time, lowering the luminosity.

When the circumstellar optical depth, $\tau_w$, becomes of order $c/v_s$ or more, the
radiation dominated shock can propagate into the circumstellar region.
In this case, substantial supernova energy can be present as internal energy at a larger radius
than would otherwise be the case, so that the effects of adiabatic expansion are lessened
and the luminosity rises.
\cite[Chevalier \& Irwin (2011)]{chevalier11} suggested that this model can describe the initial rise of 10's of days of superluminous
supernovae like SN 2006gy.
When the circumstellar interaction occurs at a large radius where diffusion of radiation
can occur rapidly and the circumstellar medium is sufficiently massive to thermalize the
supernova kinetic energy, a large fraction of the supernova energy can be radiated. 
The physical parameters, a circumstellar medium of $\sim 10~M_{\odot}$ extending out to
$\sim 10^{16}$ cm, are similar to those in the diffusion model of 
\cite[Smith \& McCray (2007)]{smith07}.
\cite[Chevalier \& Irwin (2011)]{chevalier11} used an analytic model for the emission, which has limitations; in particular, the initial
interaction was described by a self-similar solution, but that solution breaks down as
the reverse shock wave propagates into the supernova ejecta.
\cite[Ginzburg \& Balberg (2012)]{ginzburg12} undertook numerical simulations that did not have this problem and obtained comparable parameters for SN 2006gy.
In their simulations,
\cite[Moriya et al. (2013)]{moriya13} found that a density law $\rho \propto r^{-2}$ did 
not give a good fit to light curve; their favored explosion parameters are mass loss as high as $15~M_{\odot}$ and energy no more than $4\times 10^{51}$ ergs.

A general expectation of the shock breakout model is that the initial rise to maximum
luminosity should be primarily due to heating of the photosphere as the shock wave
begins to affect the stellar emission.
This effect is not clearly observed in the data on SN 2006gy, although there is some
sign of an increasing temperature in the spectra of \cite[Smith et al. (2010)]{smith10}.

A Type IIn event that apparently does show the increasing temperature in the rise
to maximum is SN 2011ht.   Early Swift observations show a clear rise in
temperature to maximum (\cite[Roming et al. 2012]{roming12}),
which is characteristic of shock breakout in the circumstellar gas.
The rise time to maximum, 40 days, is an indicator of the density of the circumstellar gas.
The supernova radiation is expected to be efficiently radiated, so that the
radiated energy is a significant fraction of the supernova energy if the energy is
thermalized by the interaction..
Interestingly, the light curve is similar in shape to that of SN 2006gy, but the absolute
magnitude is fainter by 5 magnitudes, with a radiated energy $\sim 2.5\times 10^{49}$ ergs.
The implication is that the explosion energy is smaller by a factor up to $\sim 10^2$.
The possibility that  the event was not a supernova was noted by \cite[Roming et al. (2012)]{roming12}.
However, there are supernovae that occur in this low energy range and 
\cite[Mauerhan et al. (2013)]{mauerhan13}
suggest that the late emission from SN 2011ht is due to a small amount of $^{56}$Ni.
In the end, it is not possible to definitively answer the question of whether SN 2011ht was
a supernova.

For luminous supernovae that are of Type IIn, it is very plausible that the high optical luminosity is produced by circumstellar interaction (see next section).
There are also luminous supernovae that are not of the ``n'' type.
SN 2008es is a luminous Type II (\cite[Gezari et al. 2009]{gezari09}, \cite[Miller et al. 2009]{Miller09})  and SN 2010gx is  an example of a luminous Type Ib/c (\cite[Pastorello et al. 2010]{pastorello10}).
The Type Ib/c objects outnumber those of Type II.
\cite[Chevalier \& Irwin (2011)]{chevalier11} suggested that the optical light from these objects is powered
by circumstellar interaction in which the mass loss region ends before the diffusion radius is reached by the shock front.
\cite[Ginzburg \& Balberg (2012)]{ginzburg12} carried out numerical simulations of this situation and successfully modeled the optical light curve from SN 2010gx.
As described above, an expectation in this model is that the temperature should increase in the rise to maximum luminosity.
In the case of SN  2006oz, there are observations on the rise to maximum and there is no evidence for an increasing temperature; the temperature remains roughly constant
(\cite[Leloudas et al. 2012]{leloudas12}).
Another mechanism, such as magnetar power (\cite[Kasen 
\& Bildsten 2010]{kasen10}), may be indicated.

Another superluminous event of interest is PS1-10afx (\cite[Chornock et al. 2013]{chornock13}).
In this case, the luminosity and temperature at maximum implied a photospheric radius
of $5\times 10^{15}$ cm, while the observed lines indicated an expansion velocity of 11,000 km s$^{-1}$.
The implied expansion timescale is 50 days, but the rise time for the light curve is $\sim 10-15$ days.
This evolution is naturally explained in the shock breakout view (\cite[Chevalier \& Irwin 2011]{chevalier11});
the shock front must traverse the optically thick star before shock breakout occurs.
However, the rise to luminosity maximum did not show the increasing temperature that might be expected in the breakout scenario and there is no spectral evidence for
circumstellar interaction, so the nature of the event is still in doubt.

Overall, the shock breakout scenario is attractive for the events that show
narrow line spectra because there is evidence for slowly expanding circumstellar
matter ahead of the shock front.  When these lines are not present, there is no
clear evidence for dense circumstellar matter.
There may be more than one mechanism that gives rise to very luminous supernovae.

\section{Viscous Shock Wave}

Once the forward shock wave gets into a regime where the optical depth to the
surface $\tau< c/v_s$, a radiation dominated shock can no longer be maintained and
there is a transition to a viscous shock front.
There may initially be radiative acceleration of the unshocked gas, so that a viscous shock
does not initially form, but, if the circumstellar medium is extended, the formation of
a viscous shock in inevitable.
The maximum optical depth at which it can form is somewhat less than $c/v_s$.

At low circumstellar densities the emission from the reverse shock dominates that from
the forward shock because the density is higher and the shock velocity is lower, 
$\sim 1000$ km s$^{-1}$ or less.
However, as the circumstellar density rises, the reverse shock becomes radiative first,
which leads a lower increase in luminosity with increasing density and to a dense shell that can absorb radiation from the reverse shock, thus affecting the X-ray luminosity.

The forward shock, with a velocity of 1000's of km s$^{-1}$ or more, can heat gas to a temperature
of $100v_4^2$ KeV, where $v_4^2$ is the shock velocity in units of $10^4$ km s$^{-1}$.
\cite[Katz et al. (2011)]{katz11} found that inverse Compton cooling can balance the shock heating
at a temperature of 60 keV if the optical depth is $c/v_s$ (breakout conditions) and the shock velocity
is $10^4$  km s$^{-1}$.
This estimate assumed that the electrons are heated only by Coulomb collisions with
the ions.
If there is additional collisionless heating, the gas temperature is raised.
\cite[Chevalier \& Irwin (2012)]{chevalier12} estimated regions of the shock velocity vs circumstellar density plane
where the cooling time is short compared to the age and where inverse Compton cooling is larger than bremsstralung cooling.
The regime where the cooling is rapid roughly overlaps the regime where the electron scattering optical depth
is $> 1$.
If inverse Compton losses are larger than bremsstrahlung, the X-ray emission from 
the hot shocked gas is reduced as a fraction of the shock power.

The X-rays that are emitted by the hot gas have to propagate through the preshock
circumstellar medium.
One effect of the surrounding medium is Compton recoil (\cite[Chevalier \& Irwin 2012]{chevalier12}, \cite[Svirski et al. 2012]{svirski12}), which reduces the energies
of the high energy photons to a maximum of $\sim 511/\tau_{es}^2$ keV, where
$\tau_{es}$ is the electron scattering optical depth and is assumed $>1$.  
It can be seen that $\tau_{es}>8$ is necessary to have an effect in the energy range around 10 keV that
is accessible to X-ray telescopes like {\it Chandra}.
{\it NuSTAR}, with its sensitivity to 80 keV, has a better chance of detecting this effect.

Another effect is photoabsorption by the preshock medium (\cite[Chevalier \& Irwin 2012]{chevalier12}), which has a larger effect
at low X-ray energies.
An important aspect of photoabsorption is that its occurrence requires that the absorbing
medium not be completely ionized.
For the X-ray emission that is expected from the hot gas, there is the possibility of
complete ionization (\cite[Chevalier \& Irwin 2012]{chevalier12}).
Complete ionization is expected at higher shock velocities $\sim 10^4$ km s$^{-1}$,
but not at lower velocities, $< 5000$ km s$^{-1}$.  
The result also  depends on clumping and asymmetries in
the circumstellar gas.
If the gas is not fully ionized, the column density corresponding to an electron scattering
optical depth $\tau_{es}$ is $3\times 10^{23}\tau_{es} (Z/Z_{\odot})^{-1}$ cm$^{-2}$, where $Z$ is the supernova metallicity and  $Z_{\odot}$ is the
solar metallicity.
Even for $\tau_{es}=1$, the column density is orders of magnitude larger than a typical
interstellar column density and would have a dramatic effect on X-ray emission.

A supernova that provides some test of these models is the Type IIn (narrow line) SN 2010jl.
The optical spectrum showed narrow, presumably circumstellar, emission lines
on top of broad wings that could be attributed to electron scattering (\cite[Smith et al. 2010]{smith10}).
The scattering requires an electron scattering optical depth $\ge 1$.   An
X-ray spectrum 
with the {\it Chandra} observatory at 2 months required a column density $\sim 10^{24}$ cm$^{-2}$,
(\cite[Chandra et al. 2012]{chandra12}).
At an age of 12 months, the column dropped to $\sim 3\times 10^{23}$ cm$^{-2}$, confirming that
the absorption is connected to circumstellar gas.
The observations with {\it Chandra} showed a hot thermal component with temperature $\ge 10$ keV,
corresponding to a shock velocity $\ge 3000$ km s$^{-1}$.
The emission is presumably from the forward shock front in SN 2010jl.

The presence of the large column density in SN 2010jl shows that it is possible for a large fraction of
the X-ray emission to be extinguished.
The absorption in SN 2010jl corresponds to $\tau_{es} \sim 1$;
larger values of $\tau_{es}$ are expected close to the time of shock breakout in a wind.
The expected absorption optical depth at 10 keV is $\sim 1.5\tau_{es}(Z/Z_{\odot})$ if the circumstellar region
is not fully ionized.    
\cite[Chevalier \& Irwin (2012)]{chevalier12} noted that the observed low  X-ray luminosity of SN 2006gy,
$\le 10^{40}$ ergs s$^{-1}$ (\cite[Smith et al. 2007]{smith07b}; \cite[Ofek et al. 2007]{ofek07}) compared to an optical luminosity of $3\times 10^{44}$ ergs s$^{-1}$
at a time close to maximum light (\cite[Smith et al. 2010]{smith10}), is consistent with breakout in a wind.
The case of SN 2011ht was also mentioned above as a likely case of a wind breakout.
An X-ray luminosity of $10^{39}$ ergs s$^{-1}$ was initially reported for this object based on {\it Swift} observations (\cite[Roming et al. 2012]{roming12}),
but higher spatial observations with {\it Chandra} showed that the source is not
coincident with SN 2011ht (\cite[Pooley 2012]{pooley12}).
Taking a conservative upper limit to the X-ray luminosity of $10^{39}$ ergs s$^{-1}$,
the X-ray emission is again a small fraction of the optical luminosity, which
was $3\times10^{42}$ ergs s$^{-1}$.
The strong suppression of the X-ray emission in SN 2006gy and SN 2011ht is consistent
with both of these objects being shock breakouts in a wind, although the explosion energy
is probably much smaller in SN 2011ht.

\cite[Dwarkadas  \& Gruszko (2012)]{dwarkadas12} make the point that the evolution of X-ray
emission from supernovae is typically not what is expected for a steady wind, so that estimates
of mass loss rate can be misleading.
However, the preshock density at any time can be estimated from 
$\rho_0=L/(2\pi\eta R^2v_s^3)$, where $\eta$ is the efficiency of conversion of the
shock power into the luminosity $L$.
If observations are made over a period of time, an estimate of the mass required for
the luminosity can be made and if the velocity of the mass loss can be estimated from
narrow lines, the timescale for the emission can be obtained.
The mass and timescale estimates provide useful constraints on possible mechanisms
for the mass loss.

An interesting aspect of the viscous shock front in dense mass loss is that the shock wave is expected
to be collisionless, so that diffusive shock acceleration of particles to relativistic energies
can occur (\cite[Katz et al. 2011]{katz11}, \cite[Murase et al. 2011]{murase11}).
These studies show that an unusually nearby dense interaction supernova would have
to occur to have a chance of detecting high energy gamma-ray emission from the supernova.

\section{Origin of the Dense Mass Loss}

Many of the characteristics of Type IIn supernovae require an optical depth to 
electron scattering $\ge 1$ in the circumstellar medium,
which can be expected to produce wings to narrow line profiles and obscure the inner high velocities
related to the supernova (if present).
Typical parameters for Type IIn events involve masses of $0.01-10~M_{\odot}$ 
on scales of $10^{15}-10^{16}$ cm from the supernova.
As noted above, a circumstellar mass  as high as  $10-20 ~M_{\odot}$ has been estimated for SN 2006gy.
These events have been linked to luminous blue variables (LBVs) because they
are some of the only objects known to have such extreme mass loss (\cite[Kiewe et al. 2012]{kiewe12}).
In addition, the Type IIn SN 2005gl was found to have a probable LBV progenitor in
pre-explosion images (\cite[Gal-Yam 
\& Leonard 2009]{galyam09}).
Although this link has been made, there is still the issue that the reason for
the extreme mass loss from LBVs is not understood.
In addition, stellar evolution models do not predict supernova explosions at
the time of the LBV phase.
The timing of the explosion close in time to the mass loss is a remarkable
feature of the Type IIn events.

One explanation for SN 2006gy was pulsational pair instability of massive
stars (\cite[Woosley et al. 2007]{woosley07}).
In this model, there is no terminal explosion, but the emission results from
the interaction of pulsationally driven shells.
The shells can have a mass of several $M_{\odot}$ and the radiated energy from
their interaction can be $10^{50}$ ergs.

An interesting proposal by \cite[Quataert 
\& Shiode (2012)]{quataert12} is that super-Eddington fusion reactions at the
end of the life of a massive star can generate gravity waves that deposit their
energy/momentum in the outer parts of the star, driving strong mass loss.
The prime burning phases for this are Ne and O burning, which occur in the
last year of evolution before core collapse.
However, only a fraction of massive stars show evidence for the late mass
loss and, in this hypothesis, it is unclear what determines the fraction that
show the mass loss.
A variant of this hypothesis is that the burning instability leads to expansion of
the star instead of mass loss, and the mass loss is driven by binary interaction
(\cite[Soker 2013]{soker13}).

\cite[Ofek et al. (2013)]{ofek13}  found emission from SN 2010mc in the 40 days leading up to the
explosion and suggested that the observations support the gravity wave driven
mass loss hypothesis.
The reasons are that the circumstellar velocities are relatively high,
$\sim 2000$ km s$^{-1}$ and that the brightening was observed 40 days
before explosion, when the late unstable burning phases are expected.
However, \cite[Smith et al. (2013)]{smith13} show that the light curve of SN 2010mc is very much like that
of SN 2009ip during the period June - Sept 2012, indicating that they involve
a similar physical situation.  
SN 2009ip showed eruptive
behavior over the period 2009 to 2012, somewhat longer than might be expected for the
Ne and O burning phases.
Also, high velocities observed in spectral lines are possibly the result of
electron scattering.
In any case, \cite[Fraser et al. (2013)]{fraser13} note that there is some question whether SN 2009ip was
in fact a supernova in Sept 2012, a question that can also be raised for SN 2010mc.

Another possibility for the strong mass loss is that it is driven by close binary
interaction.
Common envelope evolution is expected in close binaries, accompanied by strong
mass loss.
This has the advantage that there can be considerable variety in the events.
Also, recent observations have shown that massive stars are in close binaries
more frequently than had been thought (\cite[Sana et al. 2012]{sana12}).
The problem is to have an explosion close to the time that the mass loss occurs.
\cite[Chevalier (2012)]{chevalier12b} speculated that the event involves the spiral in of a neutron star due to
common envelope evolution.
In cases where the interaction occurs at a relatively small separation, the neutron
star spirals into the core of its companion because of the steep density gradient.
In cases where there is spiral in, but the separation is somewhat greater, the
companion star can become a red supergiant with a flat density profile, which
causes spiral in to stop.
The result is a neutron star -- helium star binary.
When the He star evolves, there is again the possibility of mass loss accompanied by the
spiral in of the neutron star to the core.

In one view, the spiral in of the neutron star to the core gives rise to a red supergiant
with a neutron star core, a Thorne-\.Zytkow star  (\cite[Thorne 
\& \.Zytkow 1977]{thorne77}).
However, neutrino losses of the matter near the neutron star may lead to strong
accretion onto the neutron star, especially when the neutron star is in the core of
the companion star  (\cite[Chevalier 1996]{chevalier96}).
The rapid accretion may lead to black hole formation and an explosion, as in the model for gamma-ray bursts (GRBs)
of \cite[Fryer \& Woosley (1998)]{fryer98}.
A variant of this model is that the rapid accretion is accompanied by strong magnetic
field amplification of the neutron star, leading to an explosion (\cite[Barkov 
\& Komissarov 2011]{barkov11}),

In this scenario for explosions preceded by strong mass loss, the explosion mechanism
is not the neutrino mechanism, which is generally favored for most core collapse
supernovae, but is related to rapid rotation and accretion on a central compact object.
Spiral in of a compact object is an efficient way of achieving rapid rotation in the
central region.
The mechanism is believed to operate for GRBs, which have been associated with
energetic ($>10^{52}$ ergs) supernovae.
Some Type IIn supernovae also appear to have had unusually large energies;
for example, SN 2003ma had an integrated bolometric luminosity of 
$4\times 10^{51}$ ergs over 4.7 years and an estimated explosion energy
$>10^{52}$ ergs  (\cite[Rest et al. 2011]{rest11}).
Studies of the neutrino mechanism for core collapse supernova explosions 
indicate that the maximum energy that can be attained is $\sim 2\times 10^{51}$ ergs
(\cite[Janka 2012]{janka12J}).
Both GRBs and some Type IIn supernovae may require an explosion mechanism involving
rotation and magnetic fields in order to produce high energy explosions.

\section{Discussion}

Although circumstellar interaction at low densities can be successfully described in
terms of spherical models, there are signs that at the higher densities considered here,
where the interaction contributes to the optical luminosity, the situation is more complex.
Galactic LBVs show a complex circumstellar structure, binary evolution can lead to
aspherical strcuture (as apparently occurred in SN 1987A), and Type IIn supernovae
often show significant polarization in their optical light.
In addition to these complexities, the Type IIn events appear to include a wide
range of stellar masses and explosion energies, indicating a variety of evolutionary paths.
We are still at the early stages of understanding these events.

\clearpage
\acknowledgement
 I am grateful to my collaborators for their participation
in this research, and to NASA grant NNX12AF90G and NSF grant AST-0807727 for support.

\begin{discussion}

\end{discussion}

\end{document}